\documentclass[12pt]{iopart}

\usepackage{amssymb}
\usepackage{graphicx}

\begin{document}

\title{On two-dimensional quasitopological field theories}

\author{P. Teotonio-Sobrinho$^1$, C. Molina$^2$ and N. Yokomizo$^1$}

\address{$^1$ Universidade de S\~ao Paulo, Instituto de F\'isica - 
DFMA \\ CP 66318, 05315-970, S\~ao Paulo - SP, Brazil}

\address{$^2$ Universidade de S\~ao Paulo, Escola de Artes, Ci\^encias e Humanidades \\
 Av. Arlindo Bettio, 1000, 03828-000, S\~ao Paulo - SP, Brazil}

\ead{teotonio@fma.if.usp.br}

\begin{abstract}
We study a class of lattice field theories in two dimensions that includes gauge theories. We show that in these theories it is possible to implement a broader notion of local symmetry, based on semi-simple Hopf algebras. A character expansion is developed for the quasitopological field theories, and partition functions are calculated with this tool. Expected values of generalized Wilson loops are defined and studied with the character expansion.
\end{abstract}


\maketitle

\section{Introduction}

Quantum field theories in low dimensional spacetimes have been extensively studied in the past decades, due to their comparative simplicity and specific applications as effective models. Among them, topological theories play an important hole, as the simplest examples of soluble systems. These models have a trivial dynamics, which depends only on the background topology. A greater understanding of topological theories in lower dimension was obtained when it was realized that they could be described with a combinatorial approach \cite{Jonsson,Bachas,Fukuma,Birmingham}.

A further step towards theories with non-trivial dynamics was attempted in \cite{Teo-98}, where quasitopological models in two dimensions were constructed on the lattice as a deformation of the topological lattice theories introduced in \cite{Fukuma}. In opposition to their topological counterparts, qua\-sitopological models have a non-trivial dynamics; yet they are still simple enough to be exactly soluble. The dynamics depends not only on the topology, but also on the area of the surface. Quasitopological theories include the usual two-dimensional Yang-Mills theories as examples, but are more general, in the sense that they describe models which cannot be expressed as gauge theories.

In the characterization of the so-called quasitopological theories, the understanding that the combinatorial description could be used to afford more than purely topological information has proved to be fundamental. The quasitopological theories are based on a formalism similar to the one used in the description of topological theories, but with weaker symmetry conditions, chosen so that non-topological data become relevant \cite{Teo-98}.

The combinatorial description leads in a natural way to an algebraic interpretation of the quasitopological theories. In this work it is shown that this algebraic formulation is compatible with additional structures which define a Hopf algebra. Hopf algebras are a natural extension of a group, offering a possible generalization for the concept of symmetry. In gauge theories, local symmetry is implemented by a gauge group. The question naturally arises of what are the local symmetries present in the more general quasitopological theories. We will show that, when the algebra describing the quasitopological model is a Hopf algebra, there is a broader notion of symmetry: the local group symmetry is replaced by a Hopf algebra local symmetry.

Hopf algebras are natural generalizations of groups. Therefore it is natural to expect they play a central role in any extension of the concept of symmetry. Quantum symmetries based on Hopf algebras have been proposed in several contexts. An example is given by the spontaneously broken Hopf symmetries in two-dimensional physical systems introduced in \cite{Bais}. Hopf algebra symmetries could also be important in usual perturbative quantum field theories \cite{Kreimer,Connes,Ebrahimi-Fard}. Hopf algebra structures underlying the AdS/CFT correspondence were studied in \cite{Janik,Plefka}.

The problem of defining a 2d lattice field theory with a Hopf algebra symmetry was previously dealt with by Buffenoir and Roche in \cite{Buffenoir}, where a solution for the case of a quasitriangular Hopf algebra symmetry was presented. There a ciliation of the lattice triangulation is imposed to solve ambiguities in the deformed gauge symmetries, and the quasitriangular character of the Hopf algebra is required for their non-commutative analogs. The same problem is investigated here with distinct methods, and a solution is described for an arbitrary triangulated compact two-dimensional surface and a semi-simple Hopf algebra structure.

Character expansions are a powerful method in the analysis of two dimensional gauge theories. In the context of the Yang-Mills theory on the plane, they were introduced by Migdal \cite{Migdal}. Kazarov used this approach in compact spaces \cite{Kazarov}. An analogous technique will be presented in the context of quasitopological field theories, and quantities of interest will be calculated with this tool, such as explicit expressions for partition functions and correlation functions.

The basic quantities in the usual Yang-Mills theories are the Wilson loop expected values, and it is reasonable to expect that they also play a relevant part in the quasitopological theories. We will generalize the Wilson loop for our context, and derive general expressions for the expected values of generalized Wilson loops in terms of character expansions.

This paper is organized as follows. In section \ref{sec:qtft-review}, we shortly review the class of topological and quasitopological theories that we are going to investigate. In section \ref{sec:non-commutative-symmetries}, we introduce the non-commutative symmetries which are implemented in the mentioned theories. In section \ref{sec:Orthogonality-relations}, orthogonality relations are derived for the algebras related to topological models. With these orthogonality relations, character expansions are computed in section \ref{sec:Character-expansions}. Expected values of Wilson loops are treated next, in section \ref{sec:Wilson-loops}. Some final comments are made in section \ref{sec:Conclusions}.

\section{Quasitopological models in two dimensions}

\label{sec:qtft-review}

Quasitopological models in the lattice were introduced and developed in \cite{Teo-98}. In order to fix notation, a short review of the theory is presented in this section. For further details we refer to \cite{Teo-98}.

The models are defined as follows. Let $T$ be a triangulation of a bidimensional compact surface $M$ of genus $g$, and let $\mathcal{I}$ be some discrete set. Consider a plaquette $P$ of $T$. A variable $\sigma_{a}$ is assigned to each link $l$ on the boundary of $P$, with $a$ taking values in $\mathcal{I}$. These are the local configurations of the theory. A Boltzman local weight $C_{ijk}(P,\epsilon)\in\mathbb{C}$ is associated to each plaquette $P$ of $T$, where the indices $i,j,k$ describe the configurations $\sigma_i,\sigma_j,\sigma_k$ at the links on its boundary, and the symbol $\epsilon$ represents possible dependences of the plaquette weights on additional parameters of interest. An invertible gluing operator $g^{ij}$ with inverse $g_{ij}$ $\left(g^{ia}g_{aj}=\delta_{j}^{i}\right)$, is attached to each link. If two faces are glued along a common link, whose configurations are $\sigma_r$ and $\sigma_s$ at these faces, then there is a gluing weight $g^{rs}(l)$ associated to such link. The partition function $Z$ of the model is defined by

\begin{equation}
Z(T,\epsilon)=\sum_{\sigma_{i_1}} \cdots \sum_{\sigma_{i_N}} \prod_{P \in T} C_{ijk}(P,\epsilon) \, \prod _{l \in T} g^{rs}(l) \, ,
\label{eq:Z-geral}
\end{equation}
where $N$ is the total number of links, the first product runs over the plaquettes of $T$, and the second product runs over the links of $T$.

The definition given in (\ref{eq:Z-geral}) encompasses a large family of lattice models. Among them are the Yang-Mills lattice gauge theories with finite groups, as shown in \cite{Teo-98}. Therefore, one may think of (\ref{eq:Z-geral}) as defining a generalization of Yang-Mills theories. Another class of examples is given by the topological models studied in \cite{Fukuma}. We will be interested in the class of quasitopological models consistent with this general form of the partition function, which were defined and studied in \cite{Teo-98}.

In the quasitopological models, the coefficients $C_{ijk}$ are such that the partition function is invariant under homeomorphisms which preserve the number of plaquettes. In other words, if $T_N$ and $T^\prime_N$ are homeomorphic bidimensional triangulations with the same number $N$ of plaquettes, then $Z(T_N,\epsilon) = Z(T^\prime_N,\epsilon)$. This symmetry holds when the weights $C_{ijk}$ satisfy the flip move relation:
\begin{equation}
C_{ij}^k(\epsilon) C_{kl}^m(\epsilon) = C_{ik}^m(\epsilon) C_{jl}^k(\epsilon)	\, ,
\label{eq:mov-flip}
\end{equation}
where the gluing operator $g^{rs}$ was used to raise indices, when upper indices appear. It may happen that for some critical value $\epsilon=\epsilon_{0}$, the weights $C_{iab}\equiv C_{iab}(\epsilon=\epsilon_{0})$ also satisfy the so-called bubble move relation:
\begin{equation}
C_{iab}C^{ba}_j = g_{ij} \, .
\label{eq:mov-bolha}
\end{equation}
In this case, the restriction to triangulations with the same number of plaquettes is removed, and any pair of homeomorphic triangulations $T$ and $T^\prime$ have the same partition function, $Z(T,\epsilon_0) = Z(T^\prime,\epsilon_0)$. Therefore, the model has topological symmetry at this critical value.

\begin{figure}
\begin{center}
	\includegraphics[scale=.9]{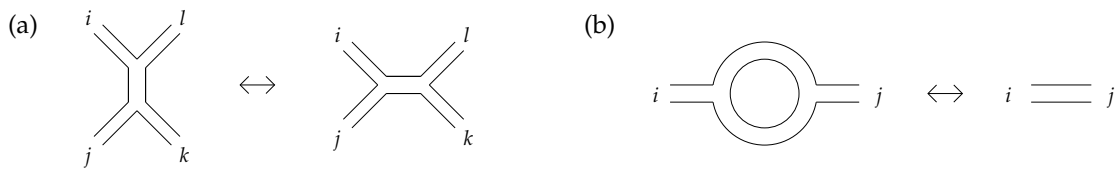}
	\end{center}
	\caption{Flip move (a) and bubble move (b), in terms of a dual graph transformation.}
	\label{fig:moves}
\end{figure}

A combinatorial interpretation of the tensorial identities (\ref{eq:mov-flip}) and (\ref{eq:mov-bolha}) is displayed in figure \ref{fig:moves}. The pictorial identities are to be understood as follows. According to the definition given in (\ref{eq:Z-geral}), the partition function $Z$ of a lattice $L$ is a tensorial expression determined by the combinatorial structure of $L$. To any transformation of the lattice there corresponds a modification of some part of the associated tensorial expression. The identities (\ref{eq:mov-flip}) and (\ref{eq:mov-bolha}) ensure that the partition function is invariant under the two Pachner moves existent in two dimensions, the flip and the bubble moves, whose action on dual graphs is displayed in figure \ref{fig:moves} (see \cite{Fukuma,Teo-98} for more details). These moves generate all homeomorphisms between triangulated bidimensional manifolds. A topological model must be invariant under both moves. Quasitopological models are characterized by the weaker condition of invariance under the flip move only.

The combinatorial definition of quasitopological models leads directly to an algebraic description. For simplicity, let $\mathcal{I}$ be a finite set with $p$ elements. Now define a vector space $A_{\epsilon}$ over $\mathbb{C}$ with basis vectors $\left\{ \phi_{1},\ldots,\phi_{p}\right\} $. For each value of $\epsilon$, an algebraic structure can be introduced on $A_{\epsilon}$ with the definition of the product
\begin{equation}
\phi_i \phi_j = C_{ij}^k(\epsilon) \phi_k \, .
\end{equation}
An immediate consequence of invariance under flip moves is that the algebras $A_{\epsilon}$ are associative,
\begin{equation}
(\phi_i \phi_j) \phi_k = \phi_i (\phi_j \phi_k ) \, , \, 
\textrm{for all values of $ \epsilon$.}
\end{equation}
Therefore, quasitopological theories are described by associative algebras. Moreover, it was proved in \cite{Fukuma} that a finite-dimensional algebra $A_{\epsilon}$ describes a topological model if and only if it is associative and semi-simple: invariance under the bubble move is equivalent to the condition of semi-simplicity.

It was shown in \cite{Teo-98} that the computation of the partition function $Z$ of a quasitopological model for any two dimensional lattice surface can be reduced to an one dimensional problem. In fact, $Z$ is given by
\begin{equation}
Z={\rm Tr} \left[ K_{\epsilon}^{(N-4g)/2} H_{\epsilon}^g \right] \, ,
\end{equation}
where $N$ and $g$ are respectively the number of triangles (plaquettes) and the genus of the triangulated surface, and the operators $K_{\epsilon}$ and $H_{\epsilon}$ are defined by its matrix elements:
\begin{eqnarray}
\left[K_\epsilon\right]_i^j = K_{ik}(\epsilon) g^{kj}, \qquad K_{ik}(\epsilon)=C_{iab}(\epsilon)C^{ba}_k(\epsilon) \, , \\
\left[H_\epsilon\right]_i^j = H_{ik}(\epsilon) g^{kj}, \qquad H_{ij}(\epsilon)=C_{ikl}(\epsilon)C^{kmn}(\epsilon)C_m^{pl}C_{npj}(\epsilon).
\end{eqnarray}
The dynamics of quasitopological field theories is also completely codified in the functional $K_{ij}(\epsilon)$: all correlation functions can be written in terms of the matrices $K_{\epsilon}^q$, where the integer $q$ depends on the number of triangles of the surface. Although the problem is greatly simplified, the explicit calculation of the partition function and correlation functions for general triangulations is still a formidable task. In later sections, further steps will be given in this direction.

The continuum limit is obtained by taking the number $N$ of plaquettes going to infinity. The parameter $\epsilon$ is interpreted as the area of each plaquette. We let $\epsilon$ approach zero with the total area $\alpha$ of the surface kept constant. Therefore $\epsilon$ and $N$ are related by $\epsilon=\alpha/N$. As we shall see, if $\epsilon=0$ is a critical point, then the model has a well defined continuum limit. At this critical point, the full topological symmetry is restored \cite{Fukuma,Teo-98}. In this case, the quasitopological models can be considered a perturbation of the topological models introduced in \cite{Fukuma}, and one finds that
\begin{equation}
\left[K_{\epsilon}^{q}\right]_a^b \longrightarrow \left[\textrm{e}^{\alpha B}\right]_a^b \, , 
\label{K_continuo}
\end{equation}
and
\begin{equation}
C_{ij}^k(\epsilon) \longrightarrow C_{ij}^k \, ,
\label{C_continuo}
\end{equation}
where $C_{ij}{}^{k}$ satisfies the relation (\ref{eq:mov-bolha}). Thus the algebra $A_0 \equiv A_{\epsilon=0}$ describes a topological model. The operator $B$ is defined by its matrix elements $B_{i}{}^{j}=B_{il}g^{lj}$, given by
\begin{equation}
B_{ij}=\left.\frac{1}{2}\frac{\partial}{\partial\epsilon}\left[C_{ikl}(\epsilon)C_i^{lk}(\epsilon)\right]\right|_{\epsilon=0}=\left.\frac{1}{2}\frac{\partial}{\partial\epsilon}K_{ij}(\epsilon)\right|_{\epsilon=0} \, ,
\end{equation}
and it follows that
\begin{equation}
B_{ij}=B_{ji} \, , \qquad C_{ki}^m B_{mj} = B_{im} C_{jk}^m \, .
\label{eq:omega-A}
\end{equation}
Moreover, any $B_{ij}$ satisfying relations (\ref{eq:omega-A}) defines a quasitopological model in the continuum limit. These conditions can be written in a more convenient form. The operator $B$ can be interpreted as a linear functional $B:A_0 \times A_0 \rightarrow\mathbb{C}$ whose action on basis elements is given by $B(\phi_i , \phi_j) = B_{ij}$. Then the conditions (\ref{eq:omega-A}) reduce to 
\begin{equation}
\begin{array}{c}
B(a,b)=B(b,a) \, , \\
B(ab,c)=B(b,ca) \, .
\end{array}
\label{eq:omega-A-forms}
\end{equation}
The set of bilinear forms which satisfy (\ref{eq:omega-A-forms}) is denoted $\Omega(A_0)$. It comes that the set of quasitopological models in the continuum limit is parametrized by a semi-simple algebra $A_0$ together with an element of $\Omega(A_0)$.

It is simple to show that $\Omega(A_0)$ is isomorphic to the space $K(A_0)$ of linear functionals over $A_0$ for which $\Phi(ab)=\Phi(ba)$, $\forall a,b \in A_0$. One explicit isomorphism $\alpha: K(A_0) \mapsto \Omega(A_0)$ is given by
\begin{equation}
\alpha(B)(a,b) = \Phi(ab) \, .
\label{eq:k-omega-iso}
\end{equation}
Therefore, it is sufficient to consider $K(A_0)$ in order to study $\Omega(A_0)$. 

In the case where $A_0 =\mathbb{C}(G)$, the group completeness and the orthogonality conditions for the group characters lead to a convenient parametrization of $K(A_0)$ and allow for the explicit calculation of $\textrm{e}^{\alpha B}$. One of the main goals of this work is to generalize these results. But before attacking this point, some additional structure will be given to the quasitopological models, in order that a generalization of the usual notion of symmetry can be introduced and explored.

\section{Hopf algebra symmetries}

\label{sec:non-commutative-symmetries}

The dynamics of quasitopological field theories is codified in a set of observables analog to the Wilson loops of lattice field theories. The symmetries of the theory are described in terms of these observables. One finds that the traditional gauge group symmetry of lattice gauge theories can be extended to a Hopf algebra symmetry acting locally on the variables of the theory.

The definition of propagators in quasitopological theories is similar to the one given by Atyiah for topological field theories in \cite{Atyiah}. A surface $T$ with boundary $\partial T = B_1 \cup B_2$, where the union is disjoint, is assumed to describe a time-evolution from the physical space $B_1$ to $B_2$. A Hilbert space $H_1$ ($H_2$) is assigned to $B_1$ ($B_2$), and to the surface $T$ there corresponds a propagator $U(T):H_1 \mapsto H_2$. In the case of quasitopological field theories, an explicit concrete description is possible, due to the available knowledge in two-dimensional topology.

Two results are central to the construction of these theories. First, there is the fact that two-dimensional manifolds (orientable and connected) with boundary are easily classified, up to homeomorphisms: they are all spherical surfaces  with some number $n$ of holes in it, and $g$ handles attached to it. Therefore, the boundary of any two-dimensional lattice is always a disjoint union of loops. Second, any such lattice can be constructed by gluing simpler pieces: three-holed spheres, sometimes called trinions, and cylinders. So one can define propagators of trinions and cylinders only, gluings being later used to construct propagators of arbitrary surfaces.

So consider a single loop with $p$ links at the boundary of some surface. We would like to assign a complex vector space to it. It was observed in the last section that a finite-dimensional algebra $A_\epsilon$ is naturally associated with any quasitopological theory, and that a basis of $A_\epsilon$ is given by $\{\phi_i,i \in \mathcal{I}\}$, where $\mathcal{I}$  indexes the local configurations of the model. We just attach an internal space $A_\epsilon$ to each link, so that the loop itself has a vector space $A_{\epsilon}^{\otimes p}$ associated with it. For a disjoint union of loops, we take the tensor product of the corresponding spaces. A natural interpretation for the vectors $\phi \in A_{\epsilon}^{\otimes p}$ is that a basis vector $\phi_{i_1} \otimes \cdots \otimes \phi_{i_p}$ describes a state in which the loop has definite configurations $\sigma_{i_1}, \dots, \sigma_{i_p}$ at its links. A general state $\phi = c_{i_1 \dots i_p} \phi_{i_1} \otimes \cdots \otimes \phi_{i_p}$ is a superposition of states with definite configurations at links.

Now a propagator must be assigned to each surface. For simplicity, consider first the case of a cylinder. The boundary of this surface is build up of two loops $l_1,l_2$. These are described by Hilbert spaces $A_{\epsilon}^{\otimes p}$ and $A_{\epsilon}^{\otimes q}$, with bases $\mathcal{B}_1 = \{\phi_{i_1} \otimes \cdots \otimes \phi_{i_p},i_l \in \mathcal{I}\}$ and $\mathcal{B}_2 = \{\phi_{j_1} \otimes \cdots \otimes \phi_{j_q},j_l \in \mathcal{I}\}$, respectively. Therefore, the surface must be described by an operator $U : A_{\epsilon}^{\otimes p} \mapsto A_{\epsilon}^{\otimes q}$. In order to define this operator, it is enough to describe its action on basis vectors. So write
\begin{equation}
U(\phi_{i_1} \otimes \cdots \otimes \phi_{i_p}) = U^{j_1 \cdots j_q}_{i_1 \cdots i_p} \phi_{j_1} \otimes \cdots \otimes \phi_{j_q} \, .
\end{equation}
and let us describe how the coefficients $U^{j_1 \cdots j_q}_{i_1 \cdots i_p}$ are calculated. First, fix configurations 
$\sigma_{i_1}, \cdots, \sigma_{i_p}$ at the links of $l_1$, and configurations $\sigma_{j_1}, \cdots, \sigma_{j_q}$ at the links
of $l_2$. Then consider an expression similar to the one employed in the definition of the partition function:
\begin{equation}
U_{i_1 \cdots i_p j_1 \cdots j_q} = \sum_\sigma \prod_{P \in T} C_{ijk}(P,\epsilon)\, \prod _{l \in T^0} g^{rs}(l) \, ,
\end{equation}
where the sum now runs over configurations on internal links only, the first product over all plaquettes, and the second product over internal links. The coefficients $U^{j_1 \cdots j_q}_{i_1 \cdots i_p}$ are defined as
\begin{equation}
U_{i_1 \cdots i_p}^{j_1 \cdots j_q} =  U_{i_1 \cdots i_p k_1 \cdots k_q} g^{k_1 j_1} \cdots g^{k_q j_q} \, .
\end{equation}
This defines the cylinder propagator. Configurations are fixed at the boundary components, which are all loops, while internal configurations are summed over. A similar procedure is used for more general surfaces. In the case of a trinion, there are three sets of indices corresponding to the three loops at its boundary, leading to coefficients $Y^{k_1 \cdots k_r}_{i_1 \cdots i_p j_1 \cdots j_q}$, which describe a propagator $Y: A_{\epsilon}^{\otimes r} \mapsto A_{\epsilon}^{\otimes p} \otimes A_{\epsilon}^{\otimes q}$. Any other propagator can be obtained from the propagators of cylinders and trinions with the interpretation of gluings of surfaces as contractions of indices in the corresponding propagators, as discussed in \cite{Teo-98}.

Explicit expressions for $U_{i_1 \cdots i_p}^{j_1 \cdots j_q}$ and $Y^{k_1 \cdots k_r}_{i_1 \cdots i_p j_1 \cdots j_q}$ were given in \cite{Teo-98}. They involve the boundary configurations only through Wilson loop variables $W(l)$ defined as 
\begin{equation}
W(l)=S_\epsilon(\phi_{i_1} \, \phi_{i_2} \cdots \phi_{i_p}) \, , 
\label{eq:loop-var}
\end{equation}
where the matrix coefficients of the operator $S: A_{\epsilon}^{\otimes p} \mapsto A_{\epsilon}^{\otimes p}$ are defined by
\begin{equation}
S^j_i = C_{iab}(\epsilon) C^{abj}(\epsilon) \, .
\end{equation}
Since propagators refer to these quantities only, Wilson loops are regarded as the physical observables of the theory.

\begin{figure}
\begin{center}
\includegraphics[scale=0.5]{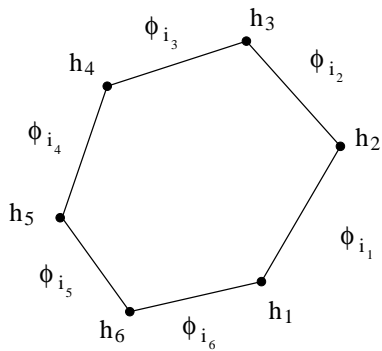}
\end{center}
\caption{Local generalized gauge transformation in a loop.}
\label{gauge-transf}
\end{figure}

Let us describe a class of transformations which preserves the propagators, up to some factors. We consider the case where $A_\epsilon$ is not only a semi-simple algebra, but a semi-simple Hopf algebra \cite{Hopf}, i.e. there is a coproduct, a counit and an antipode which satisfy all Hopf algebra axioms. Consider a particular loop, say with $p$ links, so that the loop space is $A_{\epsilon}^{\otimes p}$. The transformations are described by the algebra $\mathcal{A} \equiv A_{\epsilon}^{\otimes p}$ generated
by
\begin{eqnarray}
	h(1)_i & = & \phi_i \otimes I \otimes \cdots \otimes I \nonumber \\
 	& \vdots \nonumber \\
	h(r)_i & = & I \otimes \cdots \otimes \phi_i \otimes \cdots \otimes I \\
 	& \vdots \nonumber \\
	h(p)_i & = & I \otimes \cdots \otimes I \otimes \phi_i \, , \nonumber 
\end{eqnarray}
where $I$ is the identity. The pictorial interpretation is that $h(r)_i$ is localized at the $r-$th vertice of the loop (see figure~\ref{gauge-transf}): the action of $\mathcal{A}$ on the loop space $A_{\epsilon}^{\otimes p}$ is defined in such a manner that the generator $h(r)_{i}$ affects link variables only at links $r$ and $r+1$ (taking into account the periodicity of the link), just like gauge transformations in usual gauge theories. More specifically, the transformation $h(r)_{i}$ is defined on basis vectors of $A_{\epsilon}^{\otimes p}$ by
\begin{eqnarray}
\fl h(r)_i \left(\phi_{j_1}\otimes\cdots\otimes\phi_{j_r}\otimes\phi_{j_{r+1}}\otimes\cdots\otimes\phi_{j_{p}}\right) = \nonumber \\
=\Delta_i^{mn}\phi_{j_1}\otimes\cdots\otimes\phi_{j_r}\phi_{m}\otimes\kappa(\phi_{n})\phi_{j_{r+1}}\otimes\cdots\otimes\phi_{j_{p}} \, ,
\end{eqnarray}
where $\Delta_{i}^{mn}$ are the coefficients of the coproduct $\Delta:A_\epsilon \mapsto A_\epsilon \otimes A_\epsilon$, and $\kappa:A_\epsilon \mapsto A_\epsilon$ is the antipode of the Hopf algebra $A_\epsilon$.

The Wilson loops remain practically unchanged under these transformations, due to the Hopf algebra axioms. These variables depend only on products $\phi_{j_1} \cdots\ \phi_{j_r}$ of link variables. The factor $\phi_{j_r}\phi_{j_{r+1}}$ changes under the action of $h(r)_{i}$ according to
\begin{equation}
\phi_{j_r}\left(\Delta_i^{mn}\phi_{m}\kappa\left(\phi_{n}\right)\right)\phi_{j_{r+1}} \, ,
\label{term_to_be_mod}
\end{equation}
which simplifies to
\begin{equation}
\lambda_i \phi_{j_r} \phi_{j_{r+1}} 
\label{numerical_factor}
\end{equation}
due to the antipode relation, where $\lambda_i$ are the coefficients of the counit $\lambda: A_\epsilon \mapsto \mathbb{C}$. Thus the Wilson loop $W$ is invariant under generalized gauge transformation, up to a numerical factor. For a generic element $h\in\mathcal{A}$,
\begin{equation}
h=h^{j_1 \cdots j_p}h(1)_{j_1}\cdots h(1)_{j_p} \, ,
\end{equation}
the action on an element $\Phi\in A_{\epsilon}^{\otimes p}$ is given by
\begin{equation}
W(\Phi)\rightarrow W\left(h\Phi\right)=\lambda W(\Phi) \, ,
\label{factor_W}
\end{equation}
where $\lambda$ is the numerical factor
\begin{equation}
\lambda=\sum_{j_1 \cdots j_p} h^{j_1 \cdots j_p} \lambda_{j_1} \cdots \lambda_{j_p} \, .
\end{equation}
We recall that this factor is also present in usual gauge theories where the algebra is nothing but the group algebra (see \cite{Teo-98}). In other words, as in the case of usual gauge theories, the generalized Wilson loop is invariant under the entire group algebra only up to a factor, as given by Eq (\ref{factor_W}).

\section{Characters and orthogonality relations}

\label{sec:Orthogonality-relations}

In this section, we deal with the algebra $A_0$ of a topological model in more detail. The limit $A_0$ of $A_\epsilon$ for $\epsilon\rightarrow 0$ is the relevant algebra for computing the continuum limit. We recall that $A_\epsilon$ is semi-simple when $\epsilon=0$. In particular, we derive useful properties of the characters associated with representations of $A_0$, namely the completeness and orthogonality of the irreducible characters of the algebra. In the case where $A_0 = C(\mathbb{G})$ is the group algebra associated with a finite group $\mathbb{G}$, there are many ways of proving the orthogonality of the characters. We have found that the proof presented in \cite{James} can be adapted to the case of semi-simple algebras.

The starting point is the decomposition of the algebra in irreducible representations. It can be proved that any semi-simple algebra $A$, when considered as a $A$-module acting on itself, can be written as a direct sum
\begin{equation}
A=U_1 \oplus \cdots \oplus U_s \, ,
\label{A-decomposition}
\end{equation}
where all $U_i$ are irreducible representations of $A$. The decomposition has the following properties:
\begin{itemize}
\item every irreducible representation $U$ of $A$ is isomorphic to some $U_i$;
\item the number of $U_i$ such that $U_i \simeq U$ is equal to the dimension of $U$.
\end{itemize}
But $A$ is a finite-dimensional algebra. Since it has at least one copy of each irreducible representation, it follows that the number of irreducible representations is finite, up to isomorphisms. We denote the irreducible isomorphism classes by $R_1,\ldots,R_l$. All elements of the same class have the same dimension $d_{R_i}$ and character $\chi_{R_i}$.

Another preliminary result is that any semi-simple algebra admits an involution~$\star$ which turns it into a $C^{\star}$ algebra. That follows from Wedderburn theorem \cite{Pierse}, which states that any semi-simple algebra $A$ over the field $\mathbb{C}$ can be written as $A \simeq M_{n_1} (\mathbb{C}) \oplus \cdots \oplus M_{n_p} (\mathbb{C})$, where $M_{n_k} \left(\mathbb{C} \right)$ is the $n_k \times n_k$ matrix algebra with entries on $\mathbb{C}$. Defining the involution $\star$ as hermitian conjugation on these matrices algebras, $A$ becomes a $\star$-algebra. More structure can be introduced. The coefficients $g_{ij} = C_{iab}C^{ba}_j$ are easily computed for these matrices algebras\footnote{A convenient basis for that makes use of the decomposition $A \simeq M_{n_1} (\mathbb{C}) \oplus \cdots \oplus M_{n_p}$. Each space $M_{n_k}$ has a basis consisting of $n_ k^2$ matrices $[X_{ab}^k]_i^j = \delta_{ai} \delta_{bj}$, with $a,b =1,\ldots,n_k$. The set of all such matrices for $k=1,\ldots,p$, defines a basis of $A$.}. We use them to define a bilinear form $\langle \cdot ,\cdot \rangle : A \otimes A \mapsto \mathbb{C}$ whose action on basis elements is given by
\begin{equation}
\langle \phi_i, \phi_j \rangle = g_{ij} \, .
\end{equation}
With the involution $\star$, the sesquilinear form $\left(a,b\right)=\left\langle a^{*},b\right\rangle $ can be constructed in $A$. It can be verified that such form is a scalar product. In turn, that allows us to define a nice norm $\parallel \cdot  \parallel:A \mapsto \mathbb{R}$. Consider the operators $\rho(a): A \mapsto A$, $a \in A$, defined by $\rho(a)(v)=av, \forall v \in A$. We just put $\| a \| = \| \rho(a) \|$, where $\| \rho(a) \|$ is the norm operator. These structures collected together satisfy all $C^\star$ algebra axioms. Stating it shortly, any semi-simple algebra is a $C^\star$ algebra. In fact, the reverse is also true: all finite-dimensional $C^{*}$ algebras are isomorphic to a semi-simple algebra \cite{Murphy}, so that there exists a one-to-one correspondence between $C^\star$ algebras and topological lattice models.

The irreducible characters $\chi_{R_i}$ are elements of $\hat A$, the dual space of $A$. The metric $g^{ij}$ defines an isomorphism between $A$ and $\hat A$, and we use it to carry the scalar product from one space to the other. We let the scalar product in $\hat A$ be
\begin{equation}
\left(f,h\right)=\left(\tilde{f},\tilde{h} \right)\,\,,
\end{equation}
where $f,h \in \hat A$, $\tilde{f}=f(\phi_{i})g^{ij}\phi_{j}$ is the element in $A$ dual to $f$ (and similarly to $\tilde{h}$). With respect to this scalar product in $K(A)\subset\hat{A}$, the irreducible characters $\chi_{R_1},\ldots,\chi_{R_l}$ of $A$ can be verified to be orthogonal, $\left(\chi_{R_{i}},\chi_{R_{j}}\right)=\delta_{ij}$. Consequently, they are also linearly independent. To complete the proof, the algebra decomposition (\ref{A-decomposition}) is used to show that the dimension of $K(A)$ is equal to the number $l$ of irreducible representations. In other words $\{ \chi_{R_i} \}$ is a basis for $K(A)$.

With this result, a convenient parametrization of the space of operators $B_{ij}$ of quasitopological theories in the continuum limit can be constructed. There exists a one-to-one correspondence between such operators and elements of $K(A)$, as discussed in section \ref{sec:qtft-review}. We have just found that an arbitrary element $\phi$ of $K(A)$ can be written as $\Phi = \sum_a c_{R_a} \chi_{R_a}$, the sum running over all irreducible representations of $A$, with $c_{R_a} \in \mathbb{C}$. Using the isomorphism (\ref{eq:k-omega-iso}), we can then write any operator $B$ as $B(\phi_i,\phi_j) = \sum_a c_{R_a} \chi_{R_a} (\phi_i \phi_j)$. A similar representation is also available for $\textrm{e}^{\alpha B}$, $\alpha \in \mathbb{R}$, which can always be written as
\begin{equation}
\left[\textrm{e}^{\alpha B}\right]_a^b = \sum_{i} \textrm{e}^{\alpha B_{R_i}} d_R \chi_{R_i} \left(\phi_a \phi^b \right)\, , \qquad B_{R_i} \in \mathbb{C} \, .
\label{eq:exp-car}
\end{equation}

To conclude the discussion, let us describe a relation involving characters which will be useful in later calculations. Consider the decomposition $A=W_{1}\oplus W_{2}$, where $W_{1}$ is the direct sum of all copies of some $U_{R_i}$ and $W_{2}$ is the direct sum of the remaining components. Writing the identity element $e\in A$ as $e=e_{1}+e_{2}$, with $e_1 \in  W_1$ and $e_2 \in W_2$, it is shown that $e_1 a =d_{R_i} \chi_{R_i} (\phi^{k} a) \phi_{k}$. Now using this formula to evaluate $\chi_{R_j} (e_1 ab)$ for $j=i$ and $j \ne i$, we find that
\begin{equation}
\chi_{R_i} (\phi^k a) \chi_{R_i} (\phi_k b) = \frac{\delta_{ij}}{d_{R_i}} \chi_{R_i}(ab) \, .
\label{eq:rel_ortog}
\end{equation}
This is the main tool which will be used in the calculation of partition functions and Wilson loops in the following sections.

\section{Character expansions and partition functions}

\label{sec:Character-expansions}

We will now calculate the one-point (disk), two-point (cylinder) and  three-point (trinion) functions associated with loop configurations in $A_{\epsilon}^{\otimes p}$. All $n$-point functions can be derived from these. If a loop $\Gamma$ has $p$ vertices, the possible configurations are elements of $A_{\epsilon}^{\otimes p}$, spanned by vectors $\Phi_{\sigma}$ of the form
\begin{equation}
\Phi_{\sigma} = \phi_1 \otimes \phi_2 \otimes \cdots \otimes \phi_p \, ,
\end{equation}
where the $\phi_i$ are basis elements of $A$. The holonomy $h_{\sigma}$ associated to an element $\Phi_{\sigma}$ is
\begin{equation}
h_{\sigma} = \phi_1 \phi_2 \cdots \phi_p \in A \, .
\end{equation}
We can write $h_{\sigma}$ in terms of components as $h_{\sigma}=h_{\sigma}^j \phi_j$. Using the definition of $W\left(\Phi_{\sigma}\right)$ given in (\ref{eq:loop-var}), we see that
\begin{equation}
\left[W\left(\Phi_{\sigma}\right)\right]^a = h_{\sigma}^j S_j^a \, .
\end{equation}
On the other hand, $S_j^a$ can be expanded as
\begin{equation}
S_j^a = \sum_{i} \chi_{R_{i}} (\phi_j) \chi_{R_{i}}(\phi^a) \, ,
\end{equation}
and therefore elements of $W\left(\Phi_{\sigma}\right)$ can be written in terms of characters as
\begin{equation}
\left[ W (\Phi_{\sigma}) \right]^a = \sum_{i} \chi_{R_{i}} (h_{\sigma}) \chi_{R_{i}}(\phi^a) \, .
\end{equation}
Let us now define the related quantity $W_f:A_{\epsilon}^{\otimes p} \mapsto \mathbb{C}$, a function determined by the Wilson loops, as
\begin{equation}
W_f\left(\Phi_{\sigma}\right) = \chi_f \left(W\left(\Phi_{\sigma}\right)\right) \, ,
\label{eq:W_f}
\end{equation}
the extension to general elements of $A_{\epsilon}^{\otimes p}$ being done by linearity. That can be computed to be equal to
\begin{equation}
W_f\left(\Phi_{\sigma}\right) = \sum_{i} h_{\sigma}^j \, \chi_{R_{i}} (\phi_j) \chi_{R_{i}} (\phi^k) \chi_f (\phi_k) = \chi_f (h_{\sigma}) \, .
\end{equation}

\begin{figure}
	\begin{center}\includegraphics[scale=0.4]{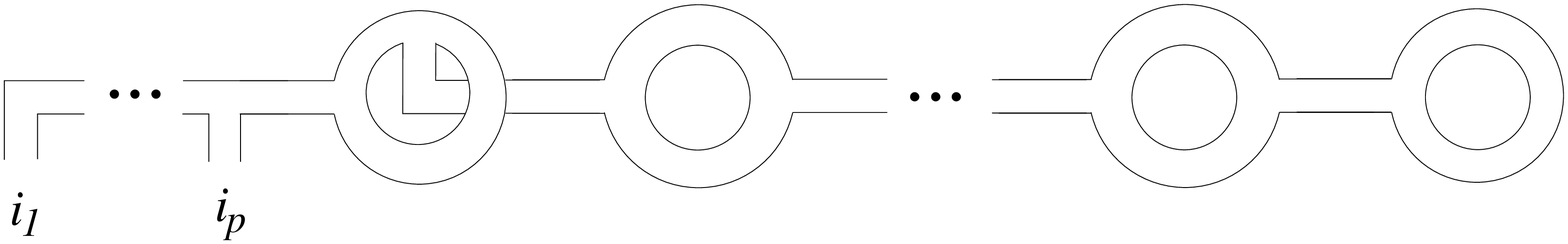}
	\end{center}
\caption{Disk diagram.}
\label{fig:disk-diagram}
\end{figure}

Using the previous results, we are now in position of calculating some quantities of interest. For a disk (one-point function) delimited by the loop $\Gamma$, the weight of a configuration $\Phi_\sigma$ is given by the diagram in figure \ref{fig:disk-diagram}, or
\begin{equation}
D\left(\Phi_{\sigma}\right)=\left[W\left(\Phi_{\sigma}\right)\right]^{a}\left[K_{\epsilon}^{q_{1}}\right]_{a}{}^{b}C_{bc}{}^{c}(\epsilon) \, ,
\end{equation}
where $q_{1}=(N-p-2)/2$, for a disk with $N$ plaquettes and $p$ links on its boundary. In the continuum limit, using (\ref{K_continuo}), (\ref{C_continuo}) and (\ref{eq:exp-car}), we can express $D\left(\Phi_{\sigma}\right)$ in terms of characters as
\begin{equation}
D\left( \Phi_{\sigma} \right) = \sum_{i} d_{R_{i}} \, \textrm{e}^{\alpha B_{R_{i}}} \chi_{R_{i}} (h_{\sigma}) \, .
\label{eq:D-car}
\end{equation}

\begin{figure}
\begin{center}
\includegraphics[scale=0.3]{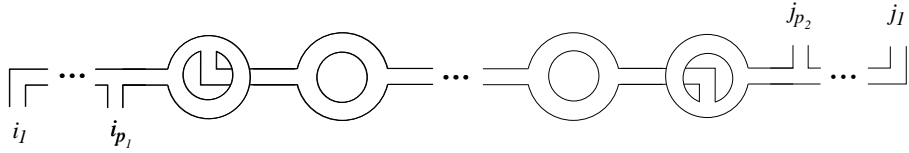}
\end{center}
\caption{Cylinder diagram.}
\label{fig:cylinder-diagram}
\end{figure}

Now consider a cylinder with $p_1$ and $p_2$ links in its boundaries. The external configurations are elements of $A_{\epsilon}^{\otimes p_1}$ and $A_{\epsilon}^{\otimes p_2}$, and the dynamics is given by the propagator $U_{\sigma_a}^{\sigma_b}:A_{\epsilon}^{\otimes p_1} \mapsto A_{\epsilon}^{\otimes p_2}$, which is written diagrammatically as shown in figure \ref{fig:cylinder-diagram}, and therefore
\begin{equation}
U\left(\Phi_{\sigma_1},\Phi_{\sigma_2}\right) = \left[W\left(\Phi_{\sigma_1}\right)\right]^a \left[K_{\epsilon}^{q_2}\right]_a^b \left[W\left(\Phi_{\sigma_2}\right)\right]_b \, ,
\end{equation}
where $q_{2}=(N-p_1-p_2-2)/2$, the triangulation being made with $N$ plaquettes and $p_1,p_2$ links at the loop boundaries. In the continuum limit,
\begin{equation}
U\left(\Phi_{\sigma_1},\Phi_{\sigma_2}\right) = \sum_{i} \textrm{e}^{\alpha B_{R_i}} \chi_{R_i}\left(h_{\sigma_1}\right) \chi_{R_i} \left(h_{\sigma_{2}}\right) \, .
\label{eq:U-car}
\end{equation}

\begin{figure}
\begin{center}
\includegraphics[scale=0.3]{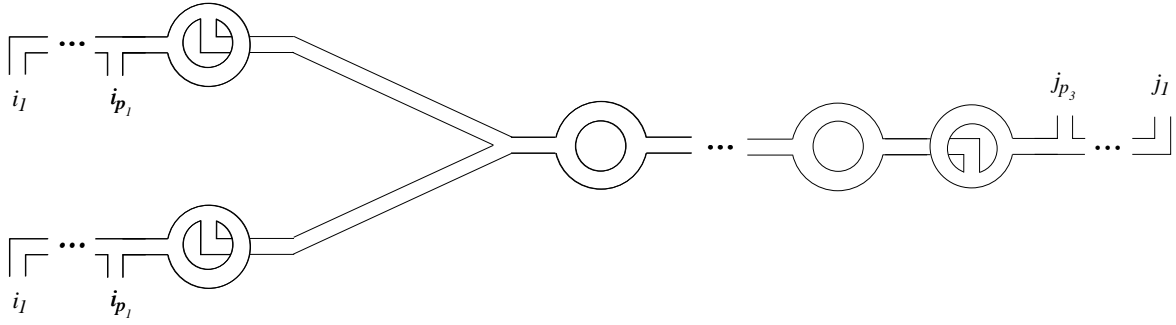}\end{center}
\caption{Diagram of the trinion.}
\label{fig:trinion-diagram}
\end{figure}

The next step is to treat the trinion (sphere with three holes), i.e. the three-point function. Its associated diagram is shown in figure \ref{fig:trinion-diagram}, and leads to
\begin{equation}
Y\left(\Phi_{\sigma_{1}},\Phi_{\sigma_{2}},\Phi^{\sigma_{3}}\right)=\left[K_{\epsilon}^{q_{3}}\right]_{a}{}^{l}C_{l}{}^{bc}(\epsilon)\left[W\left(\Phi_{\sigma_{1}}\right)\right]^{a}\left[W\left(\Phi_{\sigma_{2}}\right)\right]^{b}\left[W\left(\Phi^{\sigma_{3}}\right)\right]^{c} \, ,
\end{equation}
where $q_{3}=2(N-p_{1}-p_{2}-p_{3}-4)$, the triangulations being made with $N$ plaquettes and $p_1,p_2,p_3$ links at the loop boundaries. In the continuum limit,
\begin{equation}
Y\left(\Phi_{\sigma_1},\Phi_{\sigma_2},\Phi^{\sigma_3}\right) = \sum_{i} d_{R_i}^{-1} \textrm{e}^{\alpha B_{R_i}} \chi_{R_i} \left(h_{\sigma_1}\right) \chi_{R_i} \left(h_{\sigma_2}\right) \chi_{R_i} \left(h^{\sigma_3}\right) \, .
\label{eq:Y-car}
\end{equation}

Since any closed orientable surface is homeomorphic to the sphere or to the sphere with a finite number of handles attached, it is possible to calculate the partition function $Z$ of it in terms of the characters of $A$ using (\ref{eq:D-car}), (\ref{eq:U-car}) and (\ref{eq:Y-car}) and gluings. The partition function of the sphere, for instance, can be obtained by gluing two disks. Using (\ref{eq:D-car}) and the orthogonality relations we find that
\begin{eqnarray}
Z_{g=0} &= D\left(\Phi_{\sigma}\right)D \left(\Phi^{\sigma}\right) \nonumber \\
	&= \sum_{i \, , \, j} d_{R_i} \textrm{e}^{\alpha_1 B_{R_i}} \chi_{R_i} \left(h_{\sigma}\right)d_{R_j} e^{\alpha_2 B_{R_j}} \chi_{R_j}  \left(h^{\sigma}\right) \\
	&= \sum_i d_{R_i}^2 \textrm{e}^{\alpha B_{R_i}} \, \nonumber.
\label{Z_g0}
\end{eqnarray}
with $\alpha = \alpha_1 + \alpha_2$ being the total area of the sphere. The next step is to glue handles to the sphere, generating surfaces with genus greater than zero. Calculating the weights associated to handles and with a sphere with $n$ holes, we find the the partition function associated with a surface of genus $g$. In terms of the characters of $A$, this is given by
\begin{equation}
Z_g = \sum_i d_{R_i}^{2-2g} \textrm{e}^{\alpha B_{R_i}} \, ,
\label{eq:Z_g}
\end{equation}
and it can be seen that Eq.(\ref{eq:Z_g}) also includes the trivial topology case of genus 0.

\section{Wilson loops expected values}

\label{sec:Wilson-loops}

We will treat now more complex quantities: the expected values of Wilson loops. The first step is to find a reasonable definition for these quantities in the context of quasitopological field theories. In contrast with partition functions, whose study involved only the algebraic structure of $A$, in the case of Wilson loops expected values we have to deal with the bialgebra implicit in the Hopf algebra structure.
 
The basic requirement of a reasonable definition is that it should be basis independent. It is also desirable that, for the case of a group algebra, the definition coincides with the usual group expressions. In view of these points, a natural definition for $\left\langle W_{f}\right\rangle $ is given, in the case of a spherical surface, by
 
\begin{eqnarray}
\fl \langle W_{f} \rangle_{g=0} = \frac{1}{Z}\Delta_{a_1}^{b_1 c_1} \Delta_{a_2}^{b_2 c_2} \cdots\Delta_{a_p}^{b_p c_p} \chi_f \left(W(\phi_{b_1} \otimes \phi_{b_2} \otimes \cdots \otimes\phi_{b_p}) \right) \times \nonumber \\
\times D(\phi_{c_1} \otimes \phi_{c_2} \otimes \cdots \otimes \phi_{c_p}) D(\phi^{a_1} \otimes \phi^{a_2} \otimes \cdots \otimes \phi^{a_p}) \, .
\end{eqnarray}
This definition is explicitly independent of the choice of base in $A$. The term $1/Z$ is the usual normalization factor. 

Using previous results, the expected value $\left\langle W_{f}\right\rangle$ can be written in the continuum limit in terms of a character expansion,
\begin{equation}
\langle W_f \rangle_{g=0} = \frac{1}{Z} \sum_{i \, , \, j} d_{R_i} d_{R_j} \textrm{e}^{\alpha_1 B_{R_i} + \alpha_2 B_{R_j}} D_{R_i R_j f} \, ,
\label{W_with_D}
\end{equation}
where the $D_{R_i R_j f}$ are defined as
\begin{eqnarray}
\fl D_{R_i R_j f} = \Delta_{a_1}^{b_1 c_1} \Delta_{a_2}^{b_2 c_2} \cdots \Delta_{a_p}^{b_p c_p}  \chi_{R_i} (\phi_{c_1} \phi_{c_2} \cdots \phi_{c_p}) \chi_f (\phi_{b_1} \phi_{b_2} \cdots \phi_{b_p}) \times \nonumber \\
\times \, \chi_{R_j}(\phi^{a_1} \phi^{a_2} \cdots \phi^{a_p}) \, .
\label{def_D}
\end{eqnarray}
It is a straightforward exercise to check that, when the bialgebra is that of a group algebra, the given definition recovers the usual results \cite{Kazarov}.

It is seen from (\ref{W_with_D}) that all the new information about the coproduct is contained in the coefficients $D_{R_i R_j f}$. We will argue that these coefficients are the generalization of the Wigner coefficients for the topological algebras, and justify the necessity of a bialgebra structure in the definition of $\langle W_f \rangle_{g=0}$.

We observe that the first factors in (\ref{def_D}) can be rewritten as
\begin{eqnarray}
\fl \Delta_{a_1}^{b_1 c_1} \cdots \Delta_{a_p}^{b_p c_p} \chi_f (\phi_{b_1} \cdots \phi_{b_p}) \chi_{R_i} (\phi_{c_1} \cdots \phi_{c_p}) = \nonumber \\
= \Delta_{a_1}^{b_1 c_1} \cdots \Delta_{a_p}^{b_p c_p} \Tr \left[ (\rho_f \otimes \rho_{R_i}) (\phi_{b_1} \cdots \phi_{b_p} \otimes \phi_{c_1} \cdots \phi_{c_p})\right] \, ,
\end{eqnarray}
where $\rho_i(a)$ means the image of $a \in A$ in the irreducible representation $i$. Now from the bialgebra axiom we obtain
\begin{equation}
\fl \Delta_{a_1}^{b_1 c_1} \cdots \Delta_{a_p}^{b_p c_p} \chi_f(\phi_{b_1} \cdots \phi_{b_p}) \chi_{R_i} (\phi_{c_1} \cdots \phi_{c_p}) = \Tr \left[ (\rho_f \otimes \rho_{R_i}) \left( \Delta(h_{\sigma_a}) \right) \right] \, ,
\label{traco_do_coproduto}
\end{equation}
and from (\ref{traco_do_coproduto}), combined with the definition of the product representation for bialgebras, we have
\begin{equation}
\Delta_{a_1}^{b_1 c_1} \cdots \Delta_{a_p}^{b_p c_p} \chi_f (\phi_{b_1} \cdots \phi_{b_p}) \chi_{R_i} (\phi_{c_1} \cdots \phi_{c_p}) = \chi_{f \otimes R_i} (h_{\sigma_a}) \, ,
\label{parte_1}
\end{equation}
and therefore
\begin{equation}
D_{R_i R_j f} = \chi_{R_j}(\phi_k) \, \chi_{f \otimes R_i}(\phi^k) \, .
\label{D_prod_final}
\end{equation}
The expression (\ref{D_prod_final}) indicates that $D_{R_i R_j f}$ is a generalized version of the Wigner coefficients. These coefficients give the number of copies of the irreducible representation $R_{2}$ in the decomposition of $f \otimes  R_i$. Moreover, the product representation of two algebras is well defined only with a coproduct, justifying the necessity of a bialgebra structure in the definition (\ref{def_D}).

\subsection{Contractile Wilson loops on surfaces with non-trivial topology}

In this section we will see that the definition for the expected value of Wilson loops in a spherical surface can be generalized in a straightforward manner to surfaces with non-trivial topology. That will be made using the results (\ref{eq:D-car}), (\ref{eq:U-car}) and (\ref{eq:Y-car}) for the disk, cylinder and trinion propagators.

To simplify the manipulations, let us introduce the tensor $\tilde{\Delta}_{\sigma_{a}}{}^{\sigma_{b}\sigma_{c}}$
defined as
\begin{equation}
\tilde{\Delta}_{\sigma_a}^{\sigma_b \sigma_c} = \Delta_{a_1}^{b_1 c_1} \Delta_{a_2}^{b_2 c_2} \cdots \Delta_{a_p}^{b_p c_p} \, ,
\end{equation}
where we are using the notation $\sigma_a =(a_1, \ldots, a_p)$, $\sigma_b = (b_1, \ldots, b_p)$ and $\sigma_c = (c_1, \ldots, c_p)$. With this abbreviation, the expression for $\langle W_f \rangle_{g=0}$ can be put in a more compact form:
\begin{equation}
\langle W_f \rangle_{g=0} = \frac{1}{Z} \tilde{\Delta}_{\sigma_a}^{\sigma_b \sigma_c} [W_f]_{\sigma_b} D(\Phi_{\sigma_c}) D(\Phi^{\sigma_a}) \, ,
\end{equation}
with $[W_f]_{\sigma_b}$ given by (\ref{eq:W_f}).

\begin{figure}
\begin{center}\includegraphics[scale=0.4]{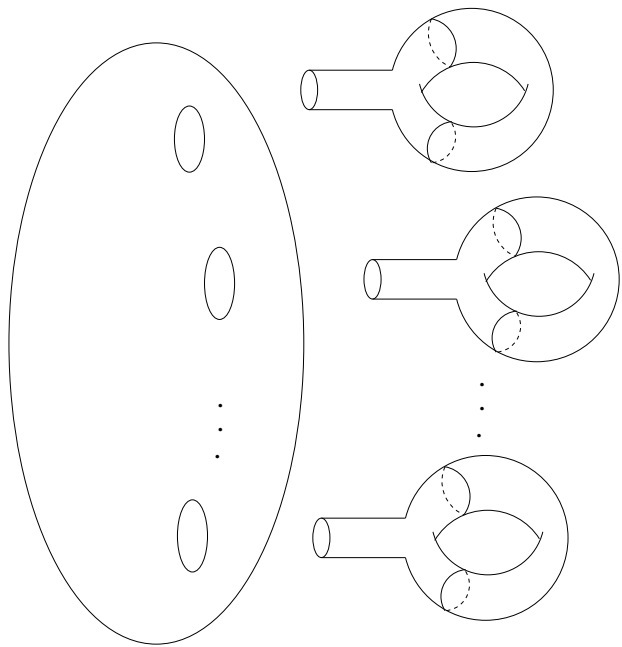}
\end{center}
\caption{Decomposition of a compact orientable surface of genus $g$.}
\label{W_genus_g}
\end{figure}

In order to define the expected value for contractile Wilson loops in a compact orientable surface of genus $g$ we use the decomposition depicted in figure \ref{W_genus_g}. The surface is build up from a spherical surface with $g$ holes where $g$ handles are attached. The contractile Wilson loop under study can be understood to lie in the spherical region, and separates it in a $(g+1)$-holed sphere and a disk. The expected value is defined from the $(g+1)$-point function $E_{g+1}$ of the cutted spherical surface, an weight $D$ of the disk bounded by the loop, and $g$ propagators $A$ corresponding to the $g$ handles, which can be constructed from the cylinder and the trinion propagators. Explicitly,
\begin{eqnarray}
\fl \langle W_f \rangle_g = \frac{1}{Z} \tilde{\Delta}_{\sigma_a}^{\sigma_b \sigma_c} [W_f]_{\sigma_b} D\left(\Phi_{\sigma_c}\right) E_{g+1}\left(\Phi^{\sigma_a},\Phi_{\sigma_{d_1}}, \ldots, \Phi_{\sigma_{d_g}}\right)  \times \nonumber \\ 
\times \, A\left(\Phi^{\sigma_{d_1}}\right) \cdots A\left( \Phi^{\sigma_{d_g}} \right) \, .
\label{eq:w-g}
\end{eqnarray}
The expression for $\left\langle W_{f}\right\rangle _{g}$ given in (\ref{eq:w-g}) can be expanded in characters, as in the case of $\langle W_f \rangle_{g=0}$. The result is
\begin{equation}
\fl \langle W_f \rangle_g = \frac{1}{Z} \sum_{i \, , \, j} d_{R_i} d_{R_j}^{1-2g} \textrm{e}^{\alpha_1 B_{R_i} + \alpha_2 B_{R_j}} \tilde{\Delta}_{\sigma_a}^{\sigma_b\sigma_c}\chi_f (h_{\sigma_b}) \chi_{R_i} (h_{\sigma_c}) \chi_{R_j} (h^{\sigma_a}) \, ,
\end{equation}
or, in terms of the coefficients $D_{R_i R_j f}$:
\begin{equation}
\langle W_f \rangle_g = \frac{1}{Z} \sum_{i \, , \, j} d_{R_i} d_{R_j}^{1-2g} \textrm{e}^{\alpha_1 B_{R_i} + \alpha_2 B_{R_j}} D_{R_i R_j f} \, .
\end{equation}
This expression generalizes the analogous results calculated in \cite{Kazarov} for the usual Yang-Mills theory in two-dimensional surfaces.

\section{Conclusions}

\label{sec:Conclusions}

Two-dimensional pure gauge theories can be considered a deformation of zero coupling gauge theories. This scenario was brought to a more general level with the introduction of quasitopological theories in \cite{Teo-98}. These are generalizations of topological lattice field theories where the dynamics is allowed to depend not only on the topology, but also on the area of the surface, inspired by what occurs in two-dimensional gauge theories. The analogy with gauge theories is complete in the case where the quasitopological algebra $A$ is a Hopf algebra. In this case, we have shown that there exists a local Hopf algebra symmetry which generalizes the usual gauge symmetry. 

Orthogonality relations and character expansion were developed for the algebras of interest in these models, and proved to be a powerful computational tool in the more general quasitopological models. They allowed for the calculation, in a straightforward manner, of partition functions associated with surfaces of arbitrary topology and area.

We have also treated expected values of Wilson loops. A definition of these quantities was missing. We observed that a natural definition, which involves Wigner coefficients as in the usual gauge theories, requires a bialgebra structure. With this definition, general expressions for the expected values in quasitopological field theories were calculated with character expansions. We obtained expressions which generalize previous results valid in the context of pure gauge theories \cite{Kazarov,Witten}.

\ack
This work was partially supported by \emph{Funda\c{c}\~{a}o de Amparo
\`{a} Pesquisa do Estado de S\~{a}o Paulo (FAPESP)} and \emph{Conselho Nacional de Desenvolvimento Cient\'{\i}fico e Tecnol\'{o}gico (CNPq)}, Brazil.

\end{document}